\documentstyle[12pt]{article}

\parskip=\medskipamount
\begin{document}

\begin{titlepage}

\begin{center}
{ \large \bf Need for the intensity-dependent pion-nucleon coupling
in multipion production processes } \\
\vspace{2.cm}
{ \normalsize \sc  M. Martinis and V. Mikuta-Martinis\footnote
{e-mail address: martinis@thphys.irb.hr

e-mail address: mikutama@thphys.irb.hr}} \\
\vspace{0.5cm}
Department of  Physics, \\
Theory Division, \\
Rudjer Bo\v skovi\' c Institute, P.O.B. 1016, \\
10000 Zagreb,CROATIA \\
\vspace{2cm}
{ \large \bf Abstract}
\end{center}
\vspace{0.5cm}
\baselineskip=24pt
We give reasons in support of the use of an effective
intensity-dependent pion-nucleon coupling Hamiltonian for describing
the properties of the pion multiplicity distribution and the
corresponding factorial moments within the thermal-density matrix
approach. We explain
the appearance of the negative-binomial (NB) distribution for pions and
the well-known empirical relation of Wr\' oblewski .
Our  model Hamiltonian is written as a linear combination
of the generators of the $SU(1,1)$ group.
We find the generating function for the pion multiplicity distribution
at finite temperature $T$ and discuss the properties of the
second-order factorial moment. Also, we show that an intensity-dependent
pion-nucleon coupling generates the squeezed states of the pion field.
At $T=0$, these squeezed states become an inherent property of the
NB distribution.

PACS number(s): 25.75.+r, 13.85.-t, 13.75

\end{titlepage}

\setcounter{page}{2}
\newpage
\baselineskip=24pt
\vspace{1.5cm}
\section{Introduction}

Recently a considerable amount of experimental information
has been accumulated on multiplicity distributions of charged
particles produced in $pp$ and $p\bar{p}$ collisions in the center-
of-mass energy range from $10 GeV$  to $1800 GeV$ .
The Koba-Nielsen-Olesen (KNO) scaling [1], which was previously
observed in the ISR c.m.energy range from 11 to 63 $GeV$  [2], was
shown to be violated in the regime of several hundred $GeV$ [3].
The violation of the KNO scaling is characterized by an
enhancement of high-multiplicity events leading to a
broadening of the multiplicity distribution with energy.

The shape of the multiplicity distribution may be described
either by its C moments,
 $C_{q} = \langle n^{q}\rangle / \langle n\rangle ^{q}$,
or by  its central moments (higher-order dispersions),
$D_{q} = \langle (n - \langle n\rangle )^{q}
\rangle ^{1/q},  q = 2,3 \ldots $.
The exact KNO scaling implies that all $C_{q}$
moments are energy independent and that all central moments $D_{q}$
satisfy a generalized Wr\' oblewski relation  [4,5] as $\langle n\rangle
\to \infty $ :
\begin{equation}
D_{q} = A_{q}\langle n\rangle  - B_{q},
\end{equation}
with the energy-independent coefficients $A_{q}$ and $B_{q}$.
For the $pp$ and $p\bar{p}$ inelastic data below $100 GeV$,
the coefficients $A_{q}$ and $B_{q}$  are approximately
equal within errors.  This fact  implies that the
elementary Poisson distribution resulting from the independent
emission of particles is ruled out.

The total multiplicity distribution $P_{n}$ of charged particles
for a wide range of energies $(22 - 900 GeV)$
is found to be well described by a negative-binomial (NB)
distribution [3,6] that belongs
to a large class of compound Poisson distributions [7].
The NB distribution is a two-step process [8] with two free parameters:
the average number of charged particles  $\langle n\rangle $
and the parameter $k$ that affects the shape (width) of the
distribution. The parameter $k$ is  related to the
dispersion $D = D_{2}$ by the relation
\begin{equation}
(\frac{D}{\langle n\rangle })^{2}  = \frac{1}{k} + \frac{1}{\langle n\rangle }.
\end{equation}
The observed broadening of the normalized multiplicity
distribution with increasing energy implies a decrease of the
parameter $k$ with energy. The KNO scaling requires constant $k$.

Although the NB distribution
gives information on the structure of correlation functions in
multiparticle production,  the question still remains
whether its clan-structure interpretation [8] is
simply a new parametrization of the data or  has a deeper physical
insight [9]. Measurements  of multiplicity distributions in $p\bar{p}$
collisions at $TeV$ energies [10] have recently shown that their shape
is clearly different from that of the NB distribution.
The distributions display the so-called medium-multiplicity
"shoulder" [11,12]. A satisfactory
explanation of this effect is still lacking [12].

In this paper we propose to study another approach to multiplicity
distributions based on  a
pion-field thermal-density operator given in terms of an effective
intensity-dependent pion-nucleon coupling Hamiltonian with $SU(1,1)$
dynamical symmetry. We assume that
the system of produced hadronic matter (pions) is in thermal
equilibrium at the temperature $T$ immediately after the collision.

The paper is organized as follows.
In Sec.2  we present the basic ideas of our  model.
In Sec.3 we discuss  the shape of multiplicity distribution,
the correlations, and  the Wr\' oblewski relation.
Finally, in Sec. 4 we  conclude with a few  remarks on the squeezing
properties of the model in connection with the possible
extension  to include  the two-pion coupling
in the effective pion-nucleon Hamiltonian.

\newpage
\section{Thermal-density operator for the pion field}

At present accelerator energies the number of secondary
particles (mostly pions) produced in hadron-hadron collisions
is large enough, so that the statistical approach to particle
production becomes reasonable . Most of the properties of
pions produced in  high-energy hadron-hadron collisions
can be expressed simply in terms of a pion-field density
operator. We neglect difficulties associated with isospin
and, for simplicity, consider the production of  single-mode pions.
As a consequence of this restriction we are only able  to
calculate multiplicity distributions and multiplicity
correlations of pions. The energy dependence of the multiplicity
distribution will reveal the scaling properties of the collision
dynamics, and the deviation of the distribution from a
Poisson distribution will reveal correlations between the
produced pions.

We expect that in high-energy collisions most of the pions
are produced in the central region
$\mid y\mid < Y$, where $Y= ln(s/m^{2})$ is the
relative rapidity of the colliding particles.
 In this region the energy - momentum
conservation has a minor effect if the transverse momenta of the
pions are limited by the dynamics.

The density operator $\hat{\rho }_{0T}$ for a free
pion field in a heat bath of  temperature $T$ is
\begin{equation}
\hat{\rho }_{0T} = \frac{1}{Z} e^{- \beta \hat{H}_{0}} ,
\,\, \beta = \frac{1}{k_{B}T},
\end{equation}
where
\begin{eqnarray}
\hat{H}_{0} & = &  \omega (a^{\dag }a + \lambda ), \\
  lnZ & = & - \beta \lambda \omega  -
ln(1 - e^{-\beta \omega }). \nonumber
\end{eqnarray}
The quantity $\lambda \omega $ in $\hat{H}_{0}$ denotes the vacuum
energy of the free pion system.  The " zero-point
energy " corresponds to $\lambda = \frac{1}{2}$.
In the limit as $T\to 0$, the density operator $\hat{\rho }_{0T}$ reduces to
$\hat{\rho }_{0} =  \mid 0\rangle \langle 0\mid $
and represents the density operator  for
the pion-field vacuum state.

The mean number of thermal (chaotic) pions is
\begin{equation}
\bar{n}_{T}  =  \frac{1}{e^{\beta \omega } - 1}.
\end{equation}

Owing to the interaction with the nucleon field
the density operator $\hat{\rho }_{0T}$ is transformed  by means
of the unitary $S$-matrix operator into
\begin{eqnarray}
\hat{\rho }_{T}  & = &  \hat{S} \hat{\rho }_{0T} \hat{S^{\dag}} \\  \nonumber
           & = &  \frac{1}{Z}e^{- \beta \hat{H}},
\end{eqnarray}
where
\begin{equation}
\hat{H} = \hat{S}\hat{H}_{0}\hat{S}^{\dag}.
\end{equation}
We regard $\hat{H}$ as an effective Hamiltonian describing
the pion-nucleon  system. At $T=0$, the $\hat{\rho }_{T}$ becomes
$\hat{\rho } = \hat{S}\mid 0\rangle \langle 0\mid \hat{S}^{\dag}$.
The knowledge of $\hat{\rho }_{T}$ gives the possibility of finding the
pion-multiplicity distribution $P_{T}(n)$ at finite temperature $T$
by calculating
\begin{equation}
P_{T}(n) = \langle n\mid \hat{\rho }_{T}\mid n\rangle  ,
\end{equation}
where $\mid n\rangle = (n!)^{-1/2}a^{\dag n}\mid 0\rangle $ are the pion
number states.

The first-order moment of $P_{T}(n)$ gives the average  multiplicity
\begin{equation}
\langle n\rangle = \sum n P_{T}(n)
\end{equation}
and higher-order moments of $P_{T}(n)$ give information on
dynamical fluctuations from $\langle n\rangle $ and also on
multipion correlations. All  higher-order moments can be
obtained from the pion-generating function
\begin{equation}
G_{T}(z) = \sum z^{n}P_{T}(n) = Tr\{\hat{\rho}_{T} z^{\hat{N}}\}
\end{equation}
by a suitable differentiation over $z$, where $\hat{N} = a^{\dag }a$.
Thus the normalized factorial moments $F_{q}$ are
\begin{equation}
F_{q} = \frac{\langle n(n-1)\ldots (n-q+1)\rangle}{\langle n\rangle ^{q}} =
\langle n\rangle ^{- q}\frac{d^{q}G_{T}(1)}{dz^{q}}
\end{equation}
and the normalized cumulant moments $K_{q}$ are
\begin{equation}
K_{q} = \langle n\rangle ^{- q}\frac{d^{q}lnG_{T}(1)}{dz^{q}}.
\end{equation}
These moments are related to each other by the formula
\begin{equation}
F_{q} = \sum_{l=0}^{q-1} {q-1 \choose l}K_{q-l}F_{l}.
\end{equation}
For the Poisson distribution, all the normalized factorial moments are
identically equal to $1$ and all  cumulants vanish for $q > 1$.

As a measure of dynamical fluctuations we shall mostly consider  the \\
$q =2$ moments. They  are directly
related to the dispersion $D$ of the multiplicity distribution $P_{T}(n)$:
\begin{equation}
F_{2} - 1 = K_{2} = (\frac{D}{\langle n\rangle })^{2} -
\frac{1}{\langle n\rangle }.
\end{equation}

The quantity $F_{2}$ is also known as the two-particle correlation
function $g^{(2)} = F_{2}$. Its numerical values are usually used to
characterize the shape of the multiplicity distribution as well as
to indicate the tendency of particles to bunch. The $g^{(2)}$ can take
the following values:
\newpage
\begin{eqnarray}
g^{(2)} & = & 1  \,\,\, (Poisson)  \nonumber   \\
      & \langle & 1 \,\,\, (sub-Poisson/antibunching) \nonumber  \\
      & \rangle & 1 \,\,\, (super-Poisson/bunching)  \\
      & = &  2 \,\,\, (chaotic) \nonumber  \\
      & \rangle & 2 \,\,\, (enhanced \,\, bunching).   \nonumber
\end{eqnarray}
The validity of the Wr\' oblewski relation means that $g^{(2)}$ takes a
value between $1$ and $2$.

The standard pion-nucleon interaction Hamiltonian is linear in the
pion field operators. In our simplified model it corresponds to the
Hamiltonian
\begin{equation}
\hat{H} = \omega (a^{\dag }a + \frac{1}{2}) + g(a^{\dag } + a ),
\end{equation}
where $\omega $ and $g$ are real and positive parameters.
This Hamiltonian
can be written in the form  (7) using the following  $S$- matrix operator:
\begin{equation}
\hat{S} = D^{\dag }( \frac{g}{\omega }) =
exp\{ \frac{g}{\omega }(a - a^{\dag })\}
\end{equation}
and $\lambda = \frac{1}{2} - (\frac{g}{\omega })^{2}$ . The density
operator $\hat{\rho }_{T}$ corresponding to  (16) describes displaced
thermal states, i.e., the superposition of coherent and thermal \\
states of pions.

The generating function $G_{T}(z)$ is [13]
\begin{equation}
G_{T}(z) = \frac{1}{1 + (1-z)\bar{n}_{T}}exp\{ - \bar{n}_{C}
\frac{1-z}{1 + (1-z)\bar{n}_{T}}\},
\end{equation}
where $\bar{n}_{C} = (g/\omega )^{2}$ denotes the average number of
coherent pions. The normalized factorial moments $F_{q}$ are
given in terms of Laguerre polynomials:
\begin{equation}
F_{q} = q!(1 + \gamma )^{-q}L_{q}(- \gamma ),
\end{equation}
where $\gamma  = \bar{n}_{C}/\bar{n}_{T}$. We note that the average
number of pions is
\begin{equation}
\langle n\rangle  =  \bar{n}_{T} + \bar{n}_{C}
\end{equation}
and the two-pion correlation function is
\begin{equation}
g^{(2)}  =   2 - (\frac{\gamma }{1 + \gamma })^{2}.
\end{equation}
According to the relation (14), the square of the dispersion $D$,
in the limit $\langle n\rangle  \to \infty $ and $\gamma  $ fixed, is
\begin{equation}
(\frac{D}{\langle n\rangle })^{2}  \to  1 - (\frac{\gamma }{1+\gamma })^{2}.
\end{equation}
It is always smaller than one as long as $\gamma \neq  0$.

On the other hand, by  keeping  the coupling
constant $g$  fixed  and letting the temperature
approach either zero or infinity ( variable $\gamma $ ),
we find that $g^{(2)} = 1$ for coherent production
and  $g^{(2)} = 2$ for chaotic production. In this case, the Wr\' oblewski
relation (1) with a constant slope $A$ cannot be obtained  .
To remedy this, we require  that $\gamma $ is an energy-independent
 parameter [14]. This requirement means
that the square of the coupling constant $g^{2}$
\begin{equation}
g^{2} = \omega ^{2}\frac{\gamma }{1 + \gamma }\langle n\rangle
\end{equation}
increases linearly with $\langle n\rangle $ or with $T$, which has the same
effect. In this way, the Wr\' oblewski relation can be satisfied
for all energies. In analogy with quantum optics [15], we shall call the
coupling of the type (23) intensity dependent.

We propose to study the following form of the effective pion-nucleon
intensity-dependent coupling Hamiltonian:
\begin{equation}
\hat{H} =  \epsilon (\hat{N} + \lambda ) +
\kappa ( a\sqrt{\hat{N} + 2\lambda  - 1} + h.c.),
\end{equation}
where $\epsilon ^{2} = \omega ^{2} + 4\kappa ^{2}$ and
$\lambda  = \langle 0\mid \hat{H}\mid 0\rangle  /\epsilon $ is related
to the vacuum energy of the system.
In the limit $\lambda \to \infty $
and $\kappa \to 0$, so that  $2\kappa ^{2}\lambda = g$ is finite,
the interaction part of the Hamiltonian $\hat{H}$  reduces to the
standard pion-nucleon Hamiltonian (16).

We observe that the operators
\begin{eqnarray}
K_{0} & = & \hat{N} + \lambda  , \\  \nonumber
K_{-} & = & a\sqrt{\hat{N} + 2\lambda - 1} ,  \\
K_{+} & = & \sqrt{\hat{N} + 2\lambda - 1}\,a^{\dag }  \nonumber
\end{eqnarray}
form the standard Holstein-Primakoff [16] realizations of the $su(1,1)$ Lie
algebra, the Casimir operator of which is
\begin{equation}
\hat{C} = K_{0}^{2} - \frac{1}{2}[K_{+}K_{-} + K_{-}K_{+}] =
\lambda (\lambda  - 1)\hat{I}.
\end{equation}
The Hamiltonian  $\hat{H}$ is thus a linear
combination of the generators of the  $SU(1,1)$ group:
\begin{equation}
\hat{H} = \epsilon  K_{0} + \kappa (K_{+} + K_{-}).
\end{equation}
The corresponding S-matrix that diagonalizes this Hamiltonian
is
\begin{equation}
\hat{S}(\theta ) = e^{ - \theta (K_{+} - K_{-})},
\end{equation}
with
\begin{equation}
th\,\theta  = \frac{2\kappa }{\epsilon }.
\end{equation}

The initial-state vector for the pion field,
$\hat{S}(\theta )\mid 0\rangle  \equiv  \mid \theta \rangle $, is [17]
\begin{equation}
\mid \theta \rangle  =  (1 - th^{2}\theta )^{\lambda }
\sum _{n}(- th\theta )^{n}\big(\frac{\Gamma (n + 2\lambda )}
{n!\Gamma (2\lambda )}\big)^{1/2}\mid n\rangle  .
\end{equation}
We note that  the matrix element squared
$\mid \langle n\mid \theta \rangle \mid ^{2}$
can be written in the form of the NB distribution
\begin{eqnarray}
p^{NB}_{n}(\theta ) & \equiv & \mid \langle n\mid \theta \rangle \mid ^{2}  \\
& = & \frac{\Gamma (n + 2\lambda )}{n!\Gamma (2\lambda )}
\big(\frac{\bar{n}(\theta )}{\bar{n}(\theta ) + 2\lambda }\big)^{n}
\big(\frac{2\lambda }{\bar{n}(\theta ) + 2\lambda }\big)^{2\lambda }, \nonumber
\end{eqnarray}
where $\bar{n}(\theta ) = 2\lambda sh^{2}(\theta )$.

The full pion-density operator
$\hat{\rho }_{T} \equiv \hat{\rho }_{T}(\theta )$ is
\begin{equation}
\hat{\rho }_{T}(\theta ) = \sum _{n} p_{T}(n)
\mid n,\theta \rangle \langle n,\theta \mid ,
\end{equation}
where $p_{T}(n) = (\bar{n}_{T})^{n}/(1+\bar{n}_{T})^{n+1}$
represents the Bose-Einstein (geometric) distribution function and
$\mid n,\theta \rangle  = \hat{S}(\theta )\mid n\rangle $.
It is easy to see that the states  $\mid n,\theta \rangle $
form a  complete orthonormal
set of eigenvectors of the Hamiltonian $\hat{H}$, i.e.,
\begin{equation}
\hat{H}\mid n,\theta \rangle   = \omega (n + \lambda )
\mid n,\theta \rangle .
\end{equation}

The calculation of the matrix element $\langle n\mid \hat{S}(\theta )
\mid m\rangle = \langle n\mid m,\theta \rangle $ is much faciliated
if we present the operator $\hat{S}(\theta )$ in the antinormal form
\begin{equation}
\hat{S}(\theta ) = e^{th(\theta )K_{-}}(ch(\theta ))^{2K_{0}}
e^{-th(\theta )K_{+}}.
\end{equation}
For $n\ge m$, we find
\begin{equation}
\langle n\mid m,\theta \rangle   = (ch(\theta ))^{-2\lambda }
(-th(\theta ))^{n-m}\big(\frac{\Gamma (n+2\lambda )m!}{\Gamma (m+2\lambda )n!}
\big)^{1/2}P_{m}^{(n-m,2\lambda - 1)}(1 - 2th^{2}(\theta )),
\end{equation}
where $P_{m}^{(\alpha ,\beta )}$ are Jacobi polynomials [18]. A similar
expression is obtained for $n\le m$:
\begin{equation}
\langle n\mid m,\theta \rangle   = (ch(\theta ))^{-2\lambda }
(th(\theta ))^{m-n}\big(\frac{\Gamma (m+2\lambda )n!}{\Gamma (n+2\lambda )m!}
\big)^{1/2}P_{n}^{(m-n,2\lambda - 1)}(1 - 2th^{2}(\theta )).
\end{equation}
These matrix elements squared are used to find the pion multiplicity
distribution
\begin{equation}
P_{T}(n) = \sum _{m}p_{T}(n)\mid \langle n\mid m,\theta \rangle \mid ^{2}
\end{equation}
and its generating function
\begin{eqnarray}
G_{T}(z) & = & \sum _{m}p_{T}(m)
\langle m,\theta \mid z^{\hat{N}}\mid m,\theta \rangle   \\
    & = & \sum _{m,n}p_{T}(m)z^{n}\mid \langle n\mid m,\theta
\rangle \mid ^{2}.  \nonumber
\end{eqnarray}

\newpage
\section{Pion-generating function and its moments }

The average multiplicity $\langle n\rangle $, the dispersion
$D$, and all  higher-order moments $\langle n^{q}\rangle $,
$q = 1,2,\ldots $, at the temperature $T$  are obtained
by a suitable differentiation over $z$ from the pion-generating
function $G_{T}(z)$.
The close analytic form of $G_{T}(z)$ can be found
by observing that
\begin{equation}
\langle m,\theta \mid z^{\hat{N}}\mid m,\theta \rangle  =
\langle \theta \mid z^{\hat{N}}\mid \theta \rangle y^{m}
P_{m}^{(0,2\lambda - 1)}(x),
\end{equation}
where
\begin{eqnarray}
x & = & \frac{z + (1-z)^{2}sh^{2}(\theta )ch^{2}(\theta )}
{z - (1-z)^{2}sh^{2}(\theta )ch^{2}(\theta )},  \\
y & = & \frac{z - (1-z)sh^{2}(\theta )}
{1 + (1-z)sh^{2}(\theta )}.  \nonumber
\end{eqnarray}
It is easy to see that
\begin{eqnarray}
\langle \theta \mid z^{\hat{N}}\mid \theta \rangle  & \equiv & G_{0}(z) \\
& = &  \sum_{m} z^{m}p^{NB}_{m}(\theta ) \nonumber  \\
& = &  [1 + (1-z)\frac{\bar{n}(\theta )}{2\lambda }]^{-2\lambda }
\end{eqnarray}
is exactly the generating function of the NB distribution
with a constant shape parameter $2\lambda $ and the average number
of pions equal to $\bar{n}(\theta )$.

Taking into account the generating function of Jacobi polynomials [18]
\begin{eqnarray}
\sum_{m}P_{m}^{(0,2\lambda -1)}(x) y^{m} & = & 2^{2\lambda  - 1}R^{-1}
(1 + y + R)^{1 - 2\lambda }, \\
R & = & \sqrt{1 -2xy + y^{2}}, \nonumber
\end{eqnarray}
we obtain the final form
of the pion-multiplicity generating function $G_{T}(z)$:
\begin{equation}
G_{T}(z) = G_{0}(z)(1 + \bar{n}_{T})^{-1}
2^{2\lambda - 1}R_{T}^{-1}(1 + y_{T}  + R_{T})^{1 - 2\lambda },
\end{equation}
where now $y_{T} = \frac{\bar{n}_{T}}{1 + \bar{n}_{T}}y$.
Using this generating function
the following average number of pions and the multiplicity dispersion
at the temperature $T$ are found:
\begin{eqnarray}
\langle n\rangle   & = & \bar{n}(\theta ) + \bar{n}_{T} +
\frac{1}{\lambda }\bar{n}(\theta )\bar{n}_{T},  \\  \nonumber
D^{2} & = & d_{T}^{2} + d^{2}(\theta )
[ 1 + \frac{2\lambda + 3}{\lambda }\bar{n}_{T} + \frac{4}{\lambda }
\bar{n}_{T}^{2}],
\end{eqnarray}
where
\begin{eqnarray}
d_{T}^{2} & = & \bar{n}_{T}^{2} + \bar{n}_{T},  \nonumber  \\
d^{2}(\theta ) & = & \frac{1}{2\lambda }\bar{n}^{2}(\theta ) +
\bar{n}(\theta ).
\end{eqnarray}

Combining the two relations in (45), we can rewrite $D^{2}$ in the
form of a quadratic function of $\langle n\rangle $:
\begin{equation}
D^{2} = A_{0}(T) + A_{1}(T)\langle n\rangle  + A_{2}(T)\langle n\rangle ^{2},
\end{equation}
with temperature-dependent coefficients $A_{i}(T), \, \, i = 0,1,2$.
In order to explain the Wr\' oblewski relation , the coefficient $A_{2}(T)$,
\begin{equation}
A_{2}(T) = 2 - \frac{6\lambda  -3}{2\lambda }(1 +
\frac{\bar{n}_{T}}{2\lambda })^{-1} + \frac{\lambda  -1}{\lambda }(1 +
\frac{\bar{n}_{T}}{2\lambda })^{-2},
\end{equation}
should be less than one. If $A_{2}(T)$ is fixed, say, by experiment, then
it gives a definite relation between the vacuum energy $\lambda $
of the pion field and the number $\bar{n}_{T}$
of the produced thermal pions. Thus, for example,the data [9]
on negatively charged particles produced in $pp$ collisions
suggest $A_{2}(T) = 1/3$. This value of
the coefficient  $A_{2}(T)$ restricts the average number of thermal
pions to $\bar{n}_{T} < 1$ for all values of $\lambda \ge  3/2$.

We notice that at  $T = 0$  the $G_{T}(z)$ becomes
the generating function of the NB distribution, $G_{0}(z)$, and
\begin{equation}
\frac{D^{2}}{\langle n\rangle ^{2}}\Big|_{T=0} =
\frac{d^{2}(\theta )}{\bar{n}^{2}(\theta )} =
\frac{1}{2\lambda } + \frac{1}{\bar{n}(\theta )},
\end{equation}
as it is to be expected from the  NB distribution .
The shape parameter $2\lambda $ of our NB distribution is given by
the vacuum energy of the pion field in the nucleon environment,
 and has nothing to do
with either the number of pion sources (cells) or the number of clans.
The Wr\' oblewski relation at $T=0$,
\begin{equation}
d(\theta ) \approx A\bar{n}(\theta ) + B, \,\,\, \bar{n}(\theta )\gg 1,
\end{equation}
has energy-independent coefficients $A = (2\lambda )^{- 1/2}$
and $B = (\lambda /2)^{1/2}$. If $\lambda  > 1/2$, we have $A < 1$.

For the temperature $T$ going to infinity, we obtain
\begin{eqnarray}
\frac{D^{2}}{\langle n\rangle ^{2}}
\Big|_{T\to\infty} & = & 2 - (1 + \frac{\bar{n}(\theta )}
{\lambda })^{- 2}   \\   \nonumber
& = &  1 + th^{2}(2\theta ).
\end{eqnarray}
This result shows that at very high temperature the distribution of
pions  will become chaotic if $\theta $ is very small.

The validity of the Wr\' oblewski relation implies that the produced
pions have a tendency to bunch. In the standard approach, the bunching
of pions is usually attributed to the presence of a quadratic (two-pion)
interaction term in the Hamiltonian (16), which is now of  the form
\begin{equation}
\hat{H} = \omega (a^{\dag }a + \frac{1}{2}) + g(a^{\dag } + a) +
\frac{1}{2}\kappa (a^{2\dag } + a^{2}).
\end{equation}
Recently the properties of the density matrix
$\hat{\rho }_{T}$, corresponding
to this Hamiltonian, were investigated for the photon field [19].
It was found that  $\hat{\rho }_{T}$ then defined the squeezed coherent
thermal states [20].

Our effective intensity-dependent coupling
Hamiltonian (24) is, however, \\
highly nonlinear
and nonquadratic in the pion field. It is ,therefore, to
be expected that it also generates squeezing that, in our model,
should depend on the value of the parameter $\lambda $.
Namely, we know that for
$\lambda \to \infty $ the pion distribution reduces
to the superposition of the coherent and thermal states
that show no squeezing .

We study the squeezing properties of the pion field in two
Hermitian quadrature operators $a_{1}$ and $a_{2}$ defined by
\begin{equation}
a = a_{1} + ia_{2},
\end{equation}
which satisfy $[a_{1},a_{2}] = i/2$. The corresponding uncertainty
relation is $\Delta a_{1}\Delta a_{2} \ge 1/4$, where variances
$\Delta a_{1,2}$ are defined by
$(\Delta a_{1,2})^{2} = \langle a_{1,2}^{2}
\rangle   - \langle a_{1,2}\rangle ^{2}$.
A state of the pion field is
considered squeezed if either $\Delta a_{1}$ or $\Delta a_{2}$ are
smaller than $1/2$. If we define the relative variance
with respect to $(\Delta a_{1,2})_{coh}^{2} = 1/4$ as
\begin{equation}
S_{1,2} = 4(\Delta a_{1,2})^{2} - 1 ,
\end{equation}
then the squeezing condition becomes
\begin{equation}
- 1 \le S_{i} < 0, \,\,\,\, i = 1 \quad \mbox{or}\quad 2.
\end{equation}
At $T=0$ our model gives
\footnote{The finite-$T$ case is considered elsewhere.}
\begin{eqnarray}
S_{1} & = & 2(\langle a^{\dag }a\rangle  + \langle a^{\dag 2}\rangle  -
2\langle a^{\dag }\rangle ^{2}),  \nonumber \\
S_{2} & = & 2(\langle a^{\dag }a\rangle - \langle a^{\dag 2}\rangle ) ,
\end{eqnarray}
where
\begin{eqnarray}
\langle a^{\dag }a\rangle  & = & \bar{n}(\theta ), \nonumber  \\
\langle a^{\dag }\rangle & = & - th(\theta )\sum_{n} p_{n}^{NB}(\theta )
\sqrt{n + 2\lambda },  \\
\langle a^{\dag 2}\rangle  & = & th^{2}(\theta )
\sum_{n} p_{n}^{NB}(\theta )
\sqrt{(n + 2\lambda )(n + 2\lambda  +1)}. \nonumber
\end{eqnarray}
It is easy to see that for $\lambda \gg 1/2$,
the relative variance is $S_{1} > 0$ and
$S_{2} \approx 0$. However, for moderate values of $\lambda > 1/2$,
and such that

$\sqrt{(n + 2\lambda )(n + 2\lambda  +1)}
\approx  (n+2\lambda ) + \frac{1}{2}- \frac{1}{8(n+2\lambda )} +
\cdots $, \\
we find squeezing for $S_{2} < 0$.

\newpage
\section{Conclusions}

In this paper we have proposed to study an
intensity-dependent pion-nucleon
coupling Hamiltonian with $SU(1,1)$  dynamical
symmetry, within
a multipion-produc\-tion model in which the  pion field is
represented by the thermal-density
operator.

We have shown that this Hamiltonian
explains in a natural way the appearance of the NB
multiplicity distribution for pions and the
Wr\' oblewski relation .The shape
parameter of the NB distribution is related to the vacuum energy
of the pion field in the nucleon environment.

Also, we have shown that, depending on the value of the parameter $\lambda $,
an intensity-dependent  pion-nucleon coupling is able to generate the
squeezed states of the pion field. At $T=0$, these squeezed states
$\mid \theta \rangle $ then become an
inherent property of the NB distribution.

For $T \neq 0$, we have found the explicit analytic form of the
pion-multiplicity generating function that may
be used for obtaining all  higher-order  moments of the pion field.

\vspace{1cm}

{ \large \bf Acknowledgment }

This work was supported by the Ministry of Science of the Republic
of Croatia under Contract No.1-03-212.

\end{document}